\documentclass[amsmath,amssymb,aps,prb,reprint,superscriptaddress,nobibnotes]{revtex4-1}

\usepackage{amsmath}
\usepackage{amsfonts}
\usepackage{amssymb}
\usepackage{graphicx}
\usepackage{upgreek}
\usepackage{color}
\usepackage{braket}
\usepackage{xr}
\makeatletter 
\renewcommand\@biblabel[1]{#1.} 
\makeatother

\definecolor{darkred}{rgb}{0.5, 0, 0}
\definecolor{darkgreen}{rgb}{0, 0.5, 0}
\definecolor{darkblue}{rgb}{0.1, 0.1, 0.7}

\usepackage[bookmarks=true,
            bookmarksnumbered=false,
            bookmarksopen=true,
            breaklinks=true,
            pdfborder={0 0 0},
            final=true,
            backref=none,
            colorlinks=true,
            linkcolor=darkblue,
            menucolor=darkblue,
            citecolor=darkblue,
            urlcolor=darkblue]{hyperref}

\newcommand{\micro}{${\upmu}$}

\newcommand{\upsub}[1]{_{\mathrm{#1}}}
\newcommand{\um}{$\,$\micro m}
\newcommand{\uev}{$\,$\micro eV}

\newcommand{\nlu}{\uev\um$^{2}$}

\begin{document}
\title{Few-photon all-optical phase rotation in a quantum-well micropillar cavity}

\author{Tintu Kuriakose}
\author{Paul M. Walker}
\email{p.m.walker@sheffield.ac.uk}
\author{Toby Dowling}
\affiliation{Department of Physics and Astronomy, University of Sheffield, S3 7RH, Sheffield, UK}
\author{Oleksandr Kyriienko}
\affiliation{Department of Physics and Astronomy, University of Exeter, EX4 4QL, Exeter, UK}
\author{Ivan A. Shelykh}
\affiliation{Science Institute, University of Iceland, Dunhagi 3, IS-107, Reykjavik, Iceland}
\affiliation{Department of Physics and Technology, ITMO University, St. Petersburg, 197101, Russia}
\author{Phillipe St-Jean}
\author{Nicola Carlon Zambon}
\author{Aristide Lema\^{i}tre}
\author{Isabelle Sagnes}
\author{Luc Legratiet}
\author{Abdelmounaim Harouri}
\author{Sylvain Ravets}
\affiliation{Centre de Nanosciences et de Nanotechnologies (C2N), Universit\'e Paris Saclay - CNRS, 911200 Palaiseau, France}
\author{M. S. Skolnick}
\affiliation{Department of Physics and Astronomy, University of Sheffield, S3 7RH, Sheffield, UK}
\affiliation{Department of Physics and Technology, ITMO University, St. Petersburg, 197101, Russia}
\author{Alberto Amo}
\affiliation{Univ. Lille, CNRS, UMR 8523 --PhLAM-- Physique des Lasers Atomes et Mol\'ecules, F-59000 Lille, France}
\author{Jacqueline Bloch}
\affiliation{Centre de Nanosciences et de Nanotechnologies (C2N), Universit\'e Paris Saclay - CNRS, 911200 Palaiseau, France}
\author{D. N. Krizhanovskii}
\affiliation{Department of Physics and Astronomy, University of Sheffield, S3 7RH, Sheffield, UK}
\affiliation{Department of Physics and Technology, ITMO University, St. Petersburg, 197101, Russia}

\maketitle
\textbf{Photonic platforms are an excellent setting for quantum technologies because weak photon-environment coupling ensures long coherence times. The second key ingredient for quantum photonics is interactions between photons, which can be provided by optical nonlinearities in the form of cross-phase-modulation (XPM). This approach underpins many proposed applications in quantum optics~\cite{Vitali2000,Munro2005,Duan2000,Glancy2008,vanEnk2001,Joo2011,Boixo2008,Simon2014,Chuang1995,Hutchinson2004,spiller2006,minzioni2019} and information processing~\cite{Brod2016,chudzicki2013}, but achieving its potential requires strong single-photon-level nonlinear phase shifts and also scalable nonlinear elements. In this work we show that the required nonlinearity can be provided by exciton-polaritons in micropillars with embedded quantum wells. These combine the strong interactions of excitons~\cite{Delteil2019,Munoz_Matutano_2019} with the scalability of micrometer-sized emitters.\cite{Amo2016,Klembt2018,Whittaker2018}. We observe XPM up to $3 \pm 1$~mrad per particle using laser beams attenuated to below single photon average intensity. With our work serving as a first stepping stone, we lay down a route for quantum information processing in polaritonic lattices.}

Quantum applications of XPM include teleportation~\cite{Vitali2000}, photon-number detection~\cite{Munro2005}, metrology~\cite{Joo2011,Boixo2008}, cryptography~\cite{Simon2014}, and quantum information processing (QIP), where it was proposed as a route to circuit-~\cite{Chuang1995} and measurement-~\cite{Hutchinson2004} based quantum computing. However, there are several challenges that need to be overcome for XPM-based photonic QIP. Frequency entanglement is known to degrade the fidelity of XPM-based quantum gates for localised modes~\cite{spiller2006,minzioni2019}. This can be overcome by cascading nonlinear resonators, with each providing moderate phase shift~\cite{Brod2016,chudzicki2013}. Such cascading naturally requires scalability of the resonators. The remaining major challenge, which we address in this paper, is to find a system with high enough single-particle XPM phase shift which is suitable for scaling.

Experimentally, XPM phase shifts of order 100-500 mrad per particle have been observed in atomic ensembles \cite{hickman2015,Venkataraman2012,firstenberg2013,beck2016} and quantum dots strongly coupled to photonic cavities~\cite{fushman2008,englund2012,bose2012}. The small size of atom-like emitters ensures strong interactions and large phase shifts but at the same time makes scalability challenging. Real atoms are not trivial to trap and manipulate while it is difficult to achieve many solid state artificial atoms with the same optical frequency and at deterministic locations on a chip.
Approaches avoiding atom-like emitters have been hindered by the small optical nonlinearity in typical Kerr media. Phase shifts from 10$^{-4}$ to 0.3 mrad per particle have been demonstrated with optical fibre and atomic vapours~\cite{Venkataraman2012,hickman2015,Matsuda2009}. Polariton micropillars, where photons are strongly coupled to excitons,~\cite{Bajoni2008,book_microcavities} are a prime candidate for combining high phase shifts and scalability. Their micrometer dimensions allow scaling into large lattices with deterministic positioning and energies identical within the linewidth~\cite{Amo2016,Klembt2018,Whittaker2018}.
Thanks to the electronic confinement in the quantum well (QW) layer, the excitonic component of polaritons provides interactions at least 1000 times larger than in weakly coupled and/or bulk semiconductors~\cite{Walker2017}.
An important feature of polariton interactions is their polarization-dependence~\cite{Renucci2005,Vladimirova2010}, which can been used to implement all-optical spin switches~\cite{Amo2010,Paraiso2010} or to break time-reversal symmetry~\cite{Bleu2017}.
Polaritonic resonators have been utilised as a source of weakly sub-Poissonian light~\cite{Delteil2019,Munoz_Matutano_2019}. However, neither XPM between distinct modes nor the polarization-dependence of interactions have been harnessed at the few particle level.
\begin{figure}
\includegraphics[width=\columnwidth]{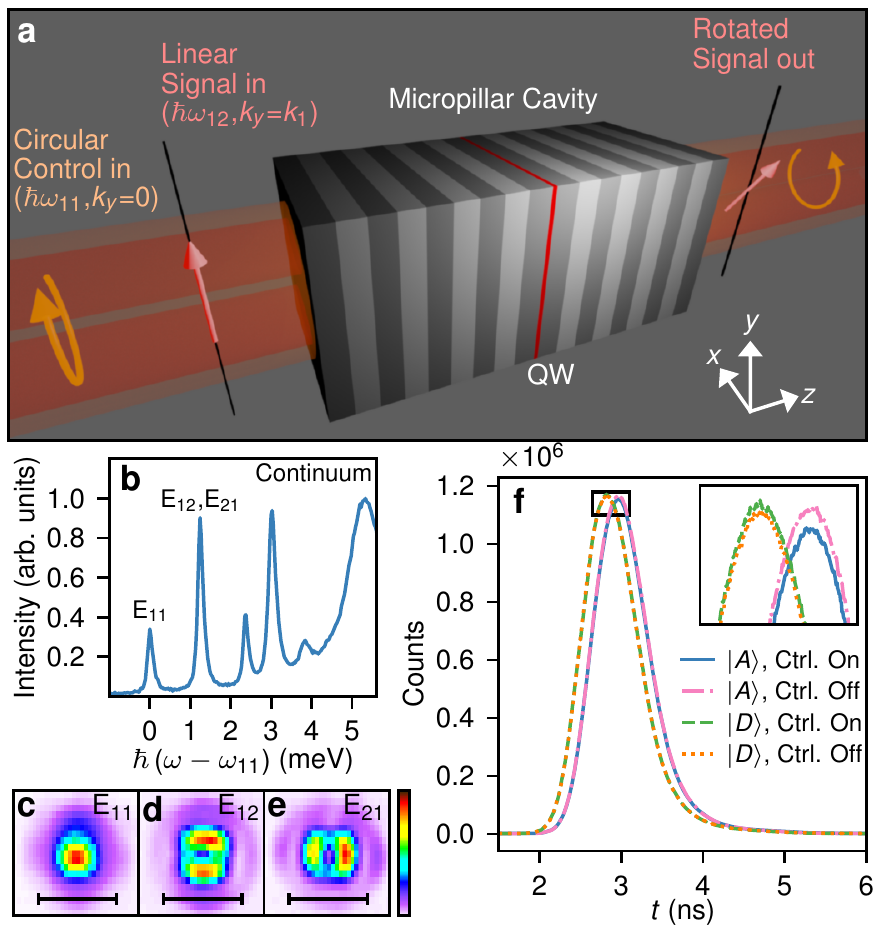}
\caption{\textbf{Sample properties.} \textbf{a,} Schematic of experimental arrangement. \textbf{b,} Photoluminescence spectrum from pillar A. $\hbar\omega_{11}=$ 1446 meV. \textbf{c-e,} Real space photoluminescence intensity maps taken at the frequencies of the ground (E$_{11}$) and first excited (E$_{12}$ and E$_{21}$) manifolds. Scale bar corresponds to 5\um{}, which is the size of the micropillar. \textbf{f,} Example of raw TCSPC curves recorded during a phase shift measurement. $\ket{D}$ and $\ket{A}$ denote the signals from APDs measuring those polarisation components while 'Ctrl. On' and 'Ctrl. Off' specify the control beam state. The inset to \textbf{f} shows a zoom of the region labelled by a black rectangle.}
\label{fig:sample}
\end{figure}

In this article we demonstrate $3$~mrad per particle XPM between two distinct optical modes, the highest amount in a device without using single emitters. As a proof of principle we demonstrate this phase rotation using a control laser attenuated to provide down to $0.13$ control particles average intensity, where the probability of more than one control polariton being in the system is less than 1\%.
We exploit the polarisation dependent interactions between polaritons~\cite{Vladimirova2010} to encode the XPM on the polarisation state of a second laser, achieving high phase sensitivity and stability.
Extrapolating our experimental results to samples with tighter photon confinement and narrower exciton linewidth~\cite{Delteil2019,Munoz_Matutano_2019}, we predict single-polariton phase shifts approaching a significant fraction of $\pi$. Using the example of XPM-based conditional-phase (CPHASE) quantum gates we show theoretically that these experimental results open new routes towards active quantum processing with exciton polaritons.

Our device is an AlGaAs air-post Fabry-Perot microcavity containing a single quantum well of the type illustrated in Fig.~\ref{fig:sample}a (see Methods~\ref{methods:sample} and Supplementary Discussion~\ref{SI:sec:expt}). All experiments were performed near to liquid helium temperature. We first characterised the micropillar using photoluminescence spectroscopy. The spectrum of discrete states resulting from the three-dimensional optical confinement can be seen in Fig.~\ref{fig:sample}b. The transverse intensity profiles of the ground ($E_{11}$) and first excited manifolds ($E_{12}$, $E_{21}$) are shown in Fig.~\ref{fig:sample}c-e. $E_{12}$ and $E_{21}$ have finite transverse wavevector in the $y$ and $x$ directions respectively, and each contain two orthogonal linear polarisation states.

The phase rotation measurement is illustrated in Fig.~\ref{fig:sample}a (see also Supplementary Discussion~\ref{SI:sec:expt}). We resonantly excited $E_{11}$ with a circularly polarized continuous-wave (CW) beam (control beam). $E_{12}$ was excited with a pulsed beam linearly polarised along the $y$ direction (signal beam). The signal linear polarisation can be decomposed into two circularly polarised components. Since polariton interactions depend strongly on relative circular polarisation~\cite{Vladimirova2010},
the presence of the control beam shifts the $E_{12}$ resonance to higher energies only for the polarisation parallel to the control beam. Consequently, the co-polarised signal component acquires a relative phase shift via XPM, resulting in a rotation of the signal beam linear polarisation angle.
Measuring this change in polarisation reveals the amount of phase shift. A quantitative analysis of the XPM and detection process is given in Supplementary Discussion~\ref{SI:sec:nonlinear_polarisation}. The overall nonlinear phase shift $\phi$ reads 
\begin{equation}\label{eq:nonlin_phase}
    \phi = \frac{2\left(g_{1}-g_{2}\right)}{\gamma/2}\left|X_{11}\right|^{2}\left|X_{12}\right|^{2}\frac{N\upsub{pol}}{A\upsub{eff}} .
\end{equation}
Here, $N\upsub{pol}$ is the mean number of control polaritons present in the cavity, $A\upsub{eff}$ is the averaged confinement area of the modes~\cite{Verger2006}, and $\gamma$ is the full width at half maximum (FWHM) signal linewidth. $\left|X_{11}\right|^{2}$ and $\left|X_{12}\right|^{2}$ are the excitonic fractions of the control and signal states. The excitonic fraction of the polaritons, and hence the strength of the interactions, increases as the polariton frequency approaches the exciton resonance. $g_{1}$ and $g_{2}$ are the interaction strengths for co- and cross-circularly-polarised excitons, respectively.
The nonlinear frequency splitting between circular-polarisation states is analogous to a Zeeman splitting caused by an effective magnetic field~\cite{Rubo2006,Larionov2010,Walker2011} and the polarisation rotation is analogous to the Faraday effect~\cite{Solnyshkov2008}.

In order to measure the XPM phase shift, we collected the light transmitted by the micropillar, filtered out the control beam using a spectrometer, and measured the diagonal ($\ket{D}$) and anti-diagonal ($\ket{A}$) signal polarisation components with the control beam chopped between on and off. Intensities were measured using time correlated single photon counting (TCSPC) allowing further separation of the pulsed signal beam from the CW control (see Methods~\ref{methods:phase} and Supplementary Discussion~\ref{SI:sec:analysis}). An example of the TCSPC data is shown in Fig.~\ref{fig:sample}f. The peaks are due to the signal pulses while effects uncorrelated with the signal pulses form a CW background which we measured using points at times far from the peak and subtracted. We then integrated the counts around the peaks to obtain total signal count rates $I_D^{(\mathrm{on})}$, $I_D^{(\mathrm{off})}$, $I_A^{(\mathrm{on})}$, $I_A^{(\mathrm{off})}$ for the $\ket{D}$ and $\ket{A}$ components with the control beam either on or off. 

Phase shift $\phi$ is deduced from the difference in polarisation degree with control beam on and off (see Supplementary Discussion~\ref{SI:sec:nonlinear_polarisation}) and is given, for small nonlinear resonance shifts compared to the linewidth, by 
\begin{equation}\label{eq:phase_from_pol_degree}
\phi\approx\left(\dfrac{I_D ^{(\mathrm{on})} - I_A ^{(\mathrm{on})}}{I_D ^{(\mathrm{on})} + I_A ^{(\mathrm{on})}}  \right ) - \left(\dfrac{I_D ^{(\mathrm{off})} - I_A ^{(\mathrm{off})}}{I_D ^{(\mathrm{off})} + I_A ^{(\mathrm{off})}} \right) .
\end{equation}
As well as measuring the phase it is important to accurately deduce the number $N\upsub{pol}$ of control polaritons in the cavity. The absolute calibration of $N\upsub{pol}$ was obtained by measurement of the cavity transfer function, carefully separating the radiative losses from other contributions to the linewidth. Such considerations lead to the relation given in Eqn.~\eqref{eq:npol} (see Methods~\ref{methods:npol} and Supplementary Discussions~\ref{SI:sec:power_meas} and \ref{SI:sec:SI_numpol}).
\begin{equation}\label{eq:npol}
E\upsub{cav} = \hbar\omega N\upsub{pol} = \frac{P\upsub{out}}{\gamma\upsub{T}\left|C_{11}\right|^{2}}
\end{equation}
Here $P\upsub{out}$ is the transmitted power and $\gamma\upsub{T}$ is the portion of the bare cavity linewidth associated with transmission through the mirror towards the detector. $\left|C_{11}\right|^{2}=1-\left|X_{11}\right|^{2}$ is the photonic fraction of the control state polaritons. We confirmed that Eqn.~\eqref{eq:npol} provides an accurate ratio of transmitted power and cavity energy by exactly solving Maxwell's equations for a wide range of cavity parameters around those of the experimental device (see Supplementary Discussion~\ref{SI:sec:SI_numpol}).
\begin{figure}
\includegraphics[width=\columnwidth]{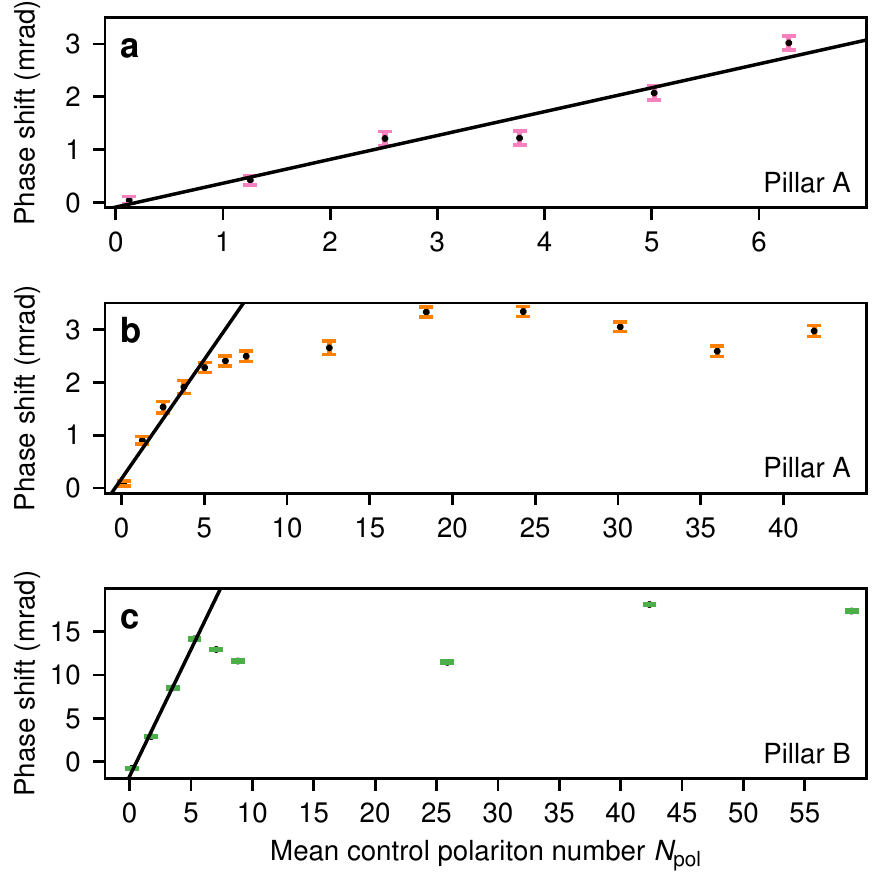}
\caption{\textbf{Measured phase shift as a function of control beam mean polariton number.} The data in \textbf{a} and \textbf{b} were measured on pillar A on two different days approximately one month apart. The data in \textbf{c} were measured on pillar B. Error bars cover the range $\pm 2\sigma$ where $\sigma$ is the standard deviation among the repeated measurements of the phase. The solid lines are best fits of straight lines passing through the origin to points with $N\upsub{pol} < 6$.}
\label{fig:phase}
\end{figure}

The phase change vs. $N\upsub{pol}$ is plotted in Fig.~\ref{fig:phase}. We considered two different micropillar cavities (A and B) with different exciton fractions and linewidths (see Methods~\ref{methods:sample}). In all three measurements in Fig.~\ref{fig:phase} the phase shift initially increases with increasing $N\upsub{pol}$. Fitting straight lines for the data with $N\upsub{pol}$ between 0.1 and 6 polaritons we deduce slopes of $0.5\pm0.2$ and $0.5\pm0.3$ mrad per polariton for the two pillar A data sets, shown in Fig.~\ref{fig:phase}a and b respectively. For pillar B, which has 7.2x larger $\left|X_{11}\right|^2\left|X_{12}\right|^2$ (see Methods~\ref{methods:sample}), the slope was $3\pm1$ mrad per polariton.

For the pillar A measurements shown in Fig.~\ref{fig:phase}b we increased $N\upsub{pol}$ up to 42. Above $\sim 6$ polaritons the phase shift saturates. A similar saturation effect was seen with pillar B (Fig.~\ref{fig:phase}c) at approximately the same occupancy. Further experiments are required to identify the mechanism behind this saturation, but it may be due to heating-induced linewidth changes~\cite{}, a partially coherent reservoir of excitons generated by the CW control beam~\cite{Sekretenko2013}, which can reduce the polarisation dependence of the interactions~\cite{Vishnevsky2012}, or to a pump-dependent change of particle statistics in the system \cite{Laussy2006}. Such effects can be overcome using pulsed, rather than CW, control excitation~\cite{Sekretenko2013,Walker2017}. We note that these effects will not be detrimental to performance since we envisage devices operating with $N\upsub{pol}\leq 1$.

Inserting our measured slopes for $\phi\left(N\upsub{pol}\right)$ in Eqn.~\eqref{eq:nonlin_phase} we find they are consistent with $g_{1}-g_{2}=11\pm4$\nlu{} and $10\pm4$\nlu{} for pillars A and B respectively. 
These are consistent with the lower end of the range established by many other groups~\cite{Amo2009,Vladimirova2009,Brichkin2011,Walker2017,Delteil2019,Munoz_Matutano_2019,Estrecho2019} indicating that we do not underestimate $N\upsub{pol}$. The agreement between pillars shows that the phase shift scales with exciton fraction as expected.
The value 3 mrad per polariton in pillar B is consistent with a blueshift of only 0.062\uev{} per polariton compared to the E$_{12}$ state linewidth 83\uev{}, which highlights the sensitivity of the technique.

\begin{figure}
\includegraphics[width=\columnwidth]{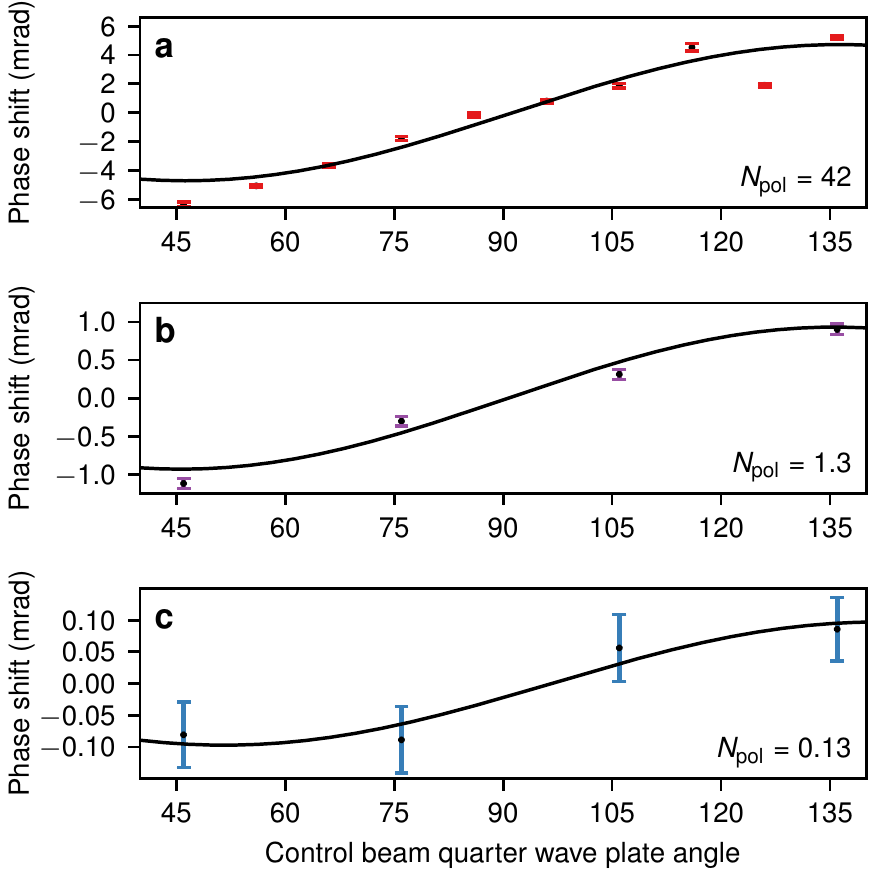}
\caption{\textbf{Phase shift dependence on control beam polarization}. \textbf{a,} $N\upsub{pol}=42\pm8$. \textbf{b,} $N\upsub{pol}=1.3\pm0.3$. \textbf{c,} $N\upsub{pol}=0.13\pm0.03$. Measurements made on pillar A. Error bars cover the range $\pm 2\sigma$ where $\sigma$ is the standard deviation among the repeated measurements of the phase.}
\label{fig:control}
\end{figure}

For our proposed Faraday-like phase rotation mechanism the induced phase should follow a sinusoidal dependence on the angle of the quarter-wave-plate (QWP) used to set the control polarisation, vanishing when the control is linearly polarised and reversing sign when it is switched to the opposite circular polarisation (see Supplementary Discussion~\ref{SI:sec:nonlinear_polarisation}). In Fig.~\ref{fig:control} we show the phase shift vs. the QWP angle for three different control beam strengths covering two orders of magnitude. The solid black lines are the best fit sinusoids. The measured phase shift agrees well with the theoretical prediction for all three control powers, reducing to zero around $90^{\circ}$ and then reversing sign. We note that the absolute magnitude of the phase shifts is different to those in Fig.~\ref{fig:phase} due to day-to-day drifts in sensitivity (see Methods~\ref{methods:noise}).

Finally, we discuss the measurements of phase at very low $N\upsub{pol}=0.13\pm0.03$. The data in Figs.~\ref{fig:phase}a and b give phase shifts of $0.04 \pm 0.06$ and $0.08 \pm 0.05$~mrad respectively. The four points in Fig.~\ref{fig:control}c also produce phase shifts larger than the uncertainty given by the error bars. We are thus able to measure a phase shift for average powers $N\upsub{pol}=0.13$, where the probability of the pillar being occupied by a single photon is 11\% and the probability of occupancy $>1$ is $<0.8\%$, based on the Poissonian statistics expected for laser fields. We are thus well inside the single photon regime.

In Table~\ref{tab:comparison} we compare the phase shifts available from various systems. Our phase shift of $3$~mrad per particle is an order of magnitude larger than in the nearest competing system which does not used atom-like emitters. While atomic and QD systems can produce larger phase shifts these can suffer from scalability challenges discussed earlier, whereas large lattices of polariton micropillars are regularly produced.
We predict that in samples with tighter photon confinement and narrower exciton linewidth,~\cite{Delteil2019} the phase shift could be two orders of magnitude larger (see Supplementary Discussion~\ref{SI:sec:fiber_phase_shift}). Moreover, an additional factor of $\sim$10 increase in interactions can also be obtained using dipolar polaritons~\cite{Rosenberg2018,Togan2018} or trion polaritons~\cite{Emmanuele2020}.
\begin{table}
\centering
\begin{tabular}{ |c|c| }
 \hline
 System & $\phi$ per particle (mrad) \\
 \hline
 Rydberg atoms in EIT regime~\cite{firstenberg2013} & 500 \\ 
 Single Cs Atom~\cite{turchette1995} & 280 \\
 Strongly coupled QD~\cite{fushman2008} & 220 \\
 \textbf{This work} & $\boldsymbol{3\pm1}$ \\
 Rb vapour in hollow core fibre~\cite{Venkataraman2012} & 0.3 \\
 Metastable Xe~\cite{hickman2015} & 0.0003 \\
 Photonic crystal fibre~\cite{Matsuda2009} & 0.0001 \\
 \hline
\end{tabular}
\caption{Comparison of XPM phase shift platforms.}
\label{tab:comparison}
\end{table}

Having demonstrated XPM phase shifts for single-polariton intensities it is interesting to consider whether XPM-based effects can be used for polaritonic QIP. Entanglement between frequency states was shown to limit the fidelity of XPM-based CPHASE gates for large phase shifts~\cite{Shapiro2006}. This obstacle can be overcome if nonlinearity is distributed over several cavities with cascaded wavepacket propagation~\cite{Shapiro2006,GeaBanacloche2010,chudzicki2013,Brod2016}. In Supplementary Discussion~\ref{SI:sec:CPHASE_gate} we analyse theoretically a potential scheme to achieve this. It requires scattering through order 10 resonators and optical circulation (suppressed backscattering). Micropillar lattices of this scale are regularly produced while directional propagation from one phase-shift element to the next can be achieved using the edge states in polariton topological insulator lattices~\cite{Klembt2018} and benefit from the rich topological physics of polaritons~\cite{Nalitov2015,Bardyn2015,Karzig2015,Solnyshkov2021}. 

As we show in Supplementary Fig.~\ref{SI:fig:fidelity} CPHASE gate fidelity depends on the size of the single-pillar phase shifts and the target phase. Full $\pi$ shift, corresponding to high-fidelity controlled-Z gate, requires large single-pillar nonlinearity-to-linewidth ratio $U\upsub{pp}/\gamma$. However even with $U\upsub{pp}/\gamma<1$ near perfect fidelity $\pi/m$ gates (integer $m$) can be achieved. When concatenated these contribute to a universal gateset and, for $m=6$, have been used in quantum hardware-based solution of optimization problems~\cite{Harrigan2021}.

While being a distant goal, we consider the ability to inject nonlinearity at the single polariton level a crucial element for many QIP protocols. In general terms, our QW polariton approach to single photon phase shifts provides the tools to optimise the balance between scalability and interaction strength for any given application.
In summary, we demonstrated a few-particle polariton XPM phase shift in a scalable on-chip platform. This opens up new approaches to a wide class of nonlinear quantum optical phenomena, and offers a route towards QIP with polaritonic lattices.

\section*{Methods and Materials}

\subsection{Sample properties}\label{methods:sample}
The sample consists of a GaAs cavity containing a single 15nm wide In$_{0.05}$Ga$_{0.95}$As QW at the electric field antinode and embedded between two Al$_{0.1}$Ga$_{0.9}$As/Al$_{0.95}$Ga$_{0.05}$As Bragg mirrors. Confinement of the light in all three dimensions results in discrete optical modes, which were measured using imaging PL spectroscopy (Fig.~\ref{fig:sample} b-e). The transverse mode profiles are similar to Hermite-Gauss modes. In the ground state manifold, labelled E$_{11}$, there are two degenerate polarisation states. In the first excited manifold the sub-manifolds E$_{12}$ and E$_{21}$, have non-zero wavevector component in the $y$ and $x$ direction respectively and each contains two orthogonal polarisation states. There is a small splitting among these four states due to a combination of TE-TM splitting and the pillars not being perfectly square. The splitting between the E$_{21}$ and E$_{12}$ sub-manifolds allowed them to be mapped separately (Fig.~\ref{fig:sample}d,e) using energies one FWHM either side of the peak (see methods~\ref{methods:PL}).

The studied sample contains many pillars with different sizes (width from 2\um{} to 5\um{}) to allow tuning of the spatial distribution of modes and their energy separation. Different detunings of pillar modes with respect to the QW exciton resonance were also available due to a wedge in the MBE-grown cavity. The experiments presented in this paper were performed on 5\um{} $\times$ 5\um{} square pillars. The square geometry allowed simple excitation of the Hermite-Gauss-like first excited state compared to more difficult beam shaping required to excite the Laguerre-Gauss-like excited states of a circular pillar. The 5\um{} size of the pillars minimised TE-TM splitting of the first excited manifold.

The sample Rabi splitting 3.4 $\pm$ 0.1 meV was obtained from a coupled exciton-photon oscillator model fit to the dispersion of pillar modes. To determine the detuning of the modes from the exciton we compared the energy splitting between the E$_{11}$ and E$_{12}$ modes with the value for very negatively detuned pillars. The splitting reduces as the photonic fraction reduces and so can be used to directly obtain the photonic (and hence excitonic) fraction of the polaritons. For pillar A the exciton fractions in the control and signal modes were $\left|X_{11}\right|^{2}\sim$ $9\%$ and $\left|X_{12}\right|^{2}\sim$ $15\%$ respectively. For pillar B they were $\left|X_{11}\right|^{2}\sim$ $25\%$ and $\left|X_{12}\right|^{2}\sim$ $42\%$.

The polariton linewidths (90\uev{} and 83\uev{} for pillars A and B respectively) were measured by monitoring the transmitted intensity of a single mode laser as it was scanned through the modes.
The measured linewidths are considerably larger than the planar cavity linewidths predicted by transfer matrix method (20-30\uev{}), which may be due to dephasing caused by the inhomogeneously broadened exciton distribution~\cite{Diniz2011}. We note that for deducing the number of polaritons we use the fraction of the linewidth due to radiative transmission towards the detector, $\gamma\upsub{T}=14\pm3$\uev{} in Eqn.~\eqref{eq:npol} (see methods~\ref{methods:npol}).

The effective mode area for nonlinear interactions $A\upsub{eff}$ is defined in Supplementary Discussion~\ref{SI:sec:nonlinear_polarisation} following the standard formula from nonlinear fiber-optics. It has the same value $A\upsub{eff}$ = 17\um{}$^{2}$ for both pillars. It was calculated using the modes of a square dielectric rod of GaAs in air obtained from the commercial eigenmode solver Lumerical MODE.

\subsection{Common experimental details}\label{methods:expt}
Experiments were performed near to liquid helium temperature. The sample was held in vacuum and mounted to a copper block connected to the heat-exchanger of a continuous flow cryostat. The copper block was held at less than 5 Kelvin, as measured using a silicon diode temperature sensor. Radiation load through the cryostat windows and the small transverse area for heat flow in a 5\um{} square micropillar may have caused the actual pillar temperatures to be higher.

The micropillars were optically excited directly (not through the substrate) using a 4 mm focal length objective (numerical aperture 0.42). Light emitted by or transmitted through the pillars was collected by a 10 mm focal length microscope objective (numerical aperture 0.6) and imaged via a set of confocal lenses onto the entrance slit of an imaging spectrometer. The spectrometer output could be switched between a CCD camera and an exit slit used to select only the signal beam for the APD measurements. The spectrometer exit slit was imaged onto the APDs via another pair of confocal lenses.

\subsection{Photoluminescence experiments}\label{methods:PL}
For the non-resonant photo-luminescence (PL) experiments excitation was with a laser at $\sim$830 nm, above the quantum well band edge, and all optical states were then populated by hot carrier relaxation. The PL spectra were recorded using a CCD camera. Mode intensity profiles were obtained by scanning the images of the modes across the spectrometer entrance slit. E$_{12}$ and E$_{21}$ were mapped separately using frequencies one FWHM (70-90\uev{}) either side of the peak. This relies on a small energy splitting between E$_{12}$ and E$_{21}$ most likely caused by slightly non-square pillars. As can be seen in Fig.~\ref{fig:sample}b the splitting was too small to resolve directly from the spectrum.

\subsection{Phase rotation measurement details}\label{methods:phase}
For the phase rotation experiments we resonantly excited the micropillar ground state with a circularly polarized control beam emitted by a CW single mode laser. At the same time, we also excited the $E_{12}$ pillar mode with a linearly polarized signal beam from a tuneable mode-locked Ti:Sapphire laser with a pulse duration of $\sim$100 ps and a repetition rate of 80 MHz. The sizes and divergences of the input control and signal beams were controlled with telescopes to match those of the pillar modes and hence optimally couple light to the microcavity. Both the signal and control beam were set to have a flat-phase beam waist of $\sim$ 3\um{} (FWHM) on the sample surface, matching the ground mode FWHM. After this, a phase mask was placed at the focus of the telescope controlling the signal beam in order to introduce a $\pi$ phase jump at the center of the signal spot on the sample surface. In this way the signal beam was converted to a Hermite-Gauss-like beam with symmetry matching that of the E$_{12}$ mode but of the wrong symmetry to excite the E$_{21}$ mode. To ensure that experimental drifts did not compromise optimal coupling the transmission of the control and signal beams was checked after every data point shown in Figs.~\ref{fig:phase} and ~\ref{fig:control} and re-optimized if necessary.

Measurement of the intensities of the two polarisation components was performed by photon counting using avalanche photodiodes (APDs) owing to their extremely small noise level. The control beam was chopped on and off at a rate of 10kHz using an electro-optic modulator driven by a square wave control signal. Counts from the APDs were sent to a time-correlated single photon counting card (TCSPC) via a router which encoded information about which APD detected the photon and whether the chopped control beam was on or off. The signal beam was attenuated so that typically 0.025 photons were detected per laser pulse on average. We avoided spurious signals in several ways. Since the control beam was CW while the signal was pulsed (see Fig.~\ref{fig:sample}f) we were able to remove any potential scattered control light reaching the APDs by subtracting the CW background from the data. This also removed any dark or APD after-pulsing counts.
The true signal counts were then obtained by summing counts around the signal peaks in the TCSPC traces. Since we measure a polarisation degree of the form shown in Eqn.~\eqref{eq:phase_from_pol_degree} any overall drifts or jitter in signal beam intensity or integration time simply cancel out. Our chopping of the control beam at $10$~kHz eliminates any control drift effects in a manner similar to lock-in detection, while collection over several minutes effectively averages out control beam jitter. Further details are given in Supplementary Discussion~\ref{SI:sec:analysis}.

\subsection{Number of polaritons}\label{methods:npol}
The number of control polaritons in the pillar $N\upsub{pol}$ was deduced using the transmitted power and the radiative loss rate through the mirror on the transmission side of the sample, $\gamma\upsub{T}$. The accuracy of Eqn.~\ref{eq:npol} was confirmed by comparison of transmitted power and stored electromagnetic energy using exact solutions of Maxwell's equations (transfer matrix method) for cavities with a wide range of parameters around those of the experimental device (see Supplementary Discussion~\ref{SI:sec:SI_numpol} for a detailed discussion).

The total radiative loss through both mirrors was obtained by measuring the linewidth $\gamma\upsub{DBR}=25\pm5$\uev{} at a very photonic detuning where the losses are dominated by the finite reflectivity of the mirrors. It agrees well with transfer matrix simulations. We then use $\gamma\upsub{T}=\eta\cdot\gamma\upsub{DBR}$ where $\eta=0.553$ is related to the relative mirror strengths and was obtained from the transfer matrix simulations.

In principle either the incident or transmitted power can be used to obtain $N\upsub{pol}$. We obtain a high transmission through the pillar with transmitted/incident power being 40\% (45\%) for the control state of pillar A (B). It is more accurate to use the transmitted power since incident power can be reflected due to imperfect mode matching.

\subsection{Statistical analysis}\label{methods:stats}
By calculating the mean and standard deviation $\sigma$ among many ($10^{3}$-$10^{4}$) repeated measurements of $\phi$, we directly obtain the average phase change and its uncertainty for each value of $N\upsub{pol}$ or quarter wave plate angle. The quoted uncertainties are $\pm 2 \sigma$ and the error bars are plotted covering the range from $-2\sigma$ to $+2\sigma$, which corresponds to the 95\% confidence interval for a normal distribution.

\subsection{Sources of noise in the data}\label{methods:noise}
There are two categories of noise contributing to the data shown in Figs.~\ref{fig:phase} and \ref{fig:control}. These are a random error in the phase of each individual data point, and systematic variations in sensitivity which occurred between individual data points and different data sets (see supplementary discussion~\ref{SI:sec:analysis}). The former arises from the Poissonian counting statistics. The latter arises because sub-linewidth changes in signal beam detuning can change the sensitivity of the measurement to the blueshift of the states. The sensitivity function is Lorentzian with the state linewidth (see Supplementary Discussion~\ref{SI:sec:nonlinear_polarisation}). During data collection small experimental drifts were corrected in-between recording each data point, resulting in small changes in sensitivity and hence some point-to-point noise. Nevertheless, the overall trends are clearly visible in the curves presented in Figs.~\ref{fig:phase} and \ref{fig:control} and they agree well with theory so we can be confident that this point-to-point noise is not too large. The dependence of sensitivity on small changes in signal laser detuning also causes small differences in scaling from one data set to another, hence the best-fit peak phase shift for $N\upsub{pol}=42$ polaritons in Fig.~\ref{fig:control}a is $5\pm1$~mrad, slightly larger than the maximum value in Fig.~\ref{fig:phase}b ($3.3\pm0.1$), which was measured on a different day.


\section*{Acknowledgements}
This work was supported by the Engineering and Physical Sciences Research Council grant EP/N031776/1, the QUANTERA project Interpol (EP/R04385X/1 and ANR-QUAN-0003-05), the Paris Ile-de-France R\'egion in the framework of DIM SIRTEQ, ERC StG ARQADIA (949730), the Marie Sk{\l}odowska-Curie individual fellowship ToPol, the H2020-FETFLAG project PhoQus (820392) and the French RENATECH network. O.~K. acknowledges the support from UK EPSRC New Investigator Award (EP/V00171X/1). A. A. acknowledges support from the Labex CEMPI (ANR-11-LABX-0007).

\section*{Author Contributions}
P. M. W. and D. N. K. conceived and designed the experiment. T. K., P. M. W. and T. D. built the experimental apparatus and performed the experiments. P. S-J, N. C.Z., A. A., S. R. and J. B. designed and characterized the sample.  A.L, IS, L.L and A.H fabricated the sample. P. M. W. analysed the data and wrote the manuscript and supplementary material with contributions from T. K. and O. K. O. K. developed the quantum theoretical description of XPM CPHASE gates. P.M.W. developed the classical theory for cavity occupancy and XPM polariton polarisation rotation. All authors contributed to discussion of the data and discussion and revision of the manuscript.

\section*{Competing Interests}
The authors declare that there are no competing interests.



\end{document}